\documentclass[sigconf,9pt]{acmart}
\renewcommand\footnotetextcopyrightpermission[1]{}
\usepackage{booktabs}
\usepackage{color}
\usepackage{graphicx}
\usepackage{url}
\usepackage{multirow}

\usepackage{xspace}

\newcommand\Mark[1]{\textsuperscript#1}
\usepackage{subfigure}
\usepackage{amsmath,chemformula}

\usepackage{pbox}
\usepackage{threeparttable,booktabs,multicol,multirow}
\setcopyright{rightsretained}
\setcopyright{none}

\begin{document}

\title{Pruning Coherent Integrated Photonic Neural Networks Using the Lottery Ticket Hypothesis%\vspace{1.2cm}
}

\author{\Large
    Sanmitra Banerjee\Mark{1}, Mahdi Nikdast\Mark{2}, Sudeep Pasricha\Mark{2}, Krishnendu Chakrabarty\Mark{1} \\
    \Mark{1}Department of Electrical and Computer Engineering, Duke University, Durham, NC, USA  \\
    \Mark{2}Department of Electrical and Computer Engineering, Colorado State University, Fort Collins, CO, USA
}

\begin{abstract}
Singular-value-decomposition-based coherent integrated photonic neural networks (SC-IPNNs) have a large footprint, suffer from high static power consumption for training and inference, and cannot be pruned using conventional DNN pruning techniques. We leverage the lottery ticket hypothesis to propose the first hardware-aware pruning method for SC-IPNNs that alleviates these challenges by minimizing the number of weight parameters. We prune a multi-layer perceptron-based SC-IPNN and show that up to 89\% of the phase angles, which correspond to weight parameters in SC-IPNNs, can be pruned with a negligible accuracy loss (smaller than 5\%) while reducing the static power consumption by up to 86\%.

%Coherent integrated photonic neural network (C-IPNN) architectures enable light-speed and ultra-low-energy accelerators for rapidly growing artificial intelligence applications. Nevertheless, C-IPNNs have a large footprint and suffer from high static power consumption for both training and inference. Pruning C-IPNNs using model compaction to reduce the number of weight parameters can potentially alleviate these problems. However, prior attempts at pruning singular-value-decomposition-based C-IPNNs (SC-IPNNs) have shown that very few parameters can be removed without significantly degrading the network accuracy. In this paper, we present the first hardware-aware pruning method for SC-IPNNs based on the lottery ticket hypothesis (LTH). We also discuss the challenges associated with pruning SC-IPNNs and show that, in addition to the classification accuracy, model-compaction techniques should be guided by a reliability assessment of the pruned networks. As a case study, we prune a multi-layer perceptron-based SC-IPNN with two hidden layers and show that up to 89\% of the phase angles, which correspond to weight parameters in SC-IPNNs, can be pruned with a negligible loss in accuracy (smaller than 5\%) while reducing the static power consumption by up to 86\%. To the best of our knowledge, this work develops the first successful and efficient pruning method for SC-IPNNs, paving the way for realizing compact and energy-efficient photonic neural networks.
\end{abstract}
\vspace{-0.5em}

%\copyrightyear{2022}
%\acmYear{2022}
%\setcopyright{acmcopyright}
%\acmConference[DAC'22]{The 56th Annual Design Automation Conference 2019}{June 2--6,2019}{Las Vegas, NV, USA}
%%\acmBooktitle{The 56th Annual Design Automation Conference 2019 (DAC'19), June 2--6, 2019, Las Vegas, NV, USA}
%\acmPrice{15.00}
%\acmISBN{978-1-4503-6725-7/19/06}
%\acmDOI{TBD}

\maketitle
\thispagestyle{plain}
\pagestyle{plain}

\section{Introduction}
Silicon photonics can enable compact, ultra-fast, and ultra-low-energy artificial intelligence (AI) accelerators, realizing a promising framework for the emerging class of information processing machines \cite{sunny2021survey}. In particular, leveraging the inherently parallel nature and high-speed of optical-domain computation, coherent integrated photonic neural networks (C-IPNNs), which operate with a single wavelength, can reduce the computational complexity of matrix multiplication---the most compute-intensive operation in deep neural networks (DNNs)---from $O(N^2)$ to $O(1)$ \cite{cheng2020silicon}. Using singular value decomposition (SVD), several C-IPNNs (referred to as SC-IPNNs in this paper) have been recently proposed \cite{sunny2021survey}. Compared to noncoherent photonic neural networks that use multiple wavelengths, SC-IPNNs are more resilient to inter-channel crosstalk and do not require power-hungry wavelength-conversion steps \cite{cheng2020silicon}.  

%Using singular value decomposition (SVD), the weight matrices in the linear layers in multi-layer perceptrons (MLPs) are factorized into two unitary and one diagonal matrix; these matrices are realized using an array of Mach-Zehnder interferometers (MZIs). Each MZI consists of two phase-shifters (PhS) with tunable phase angles and two 50:50 beam splitters (BeS). Prior work shows that $N\timesN$ unitary and diagonal matrices can be represented by tuning the phase angles in an array of $N(N-1)/2$ and $N$ MZIs respectively \cite{clements2016optimal}.  

\begin{figure}[t]
 \centering
  \includegraphics[width=.46\textwidth]{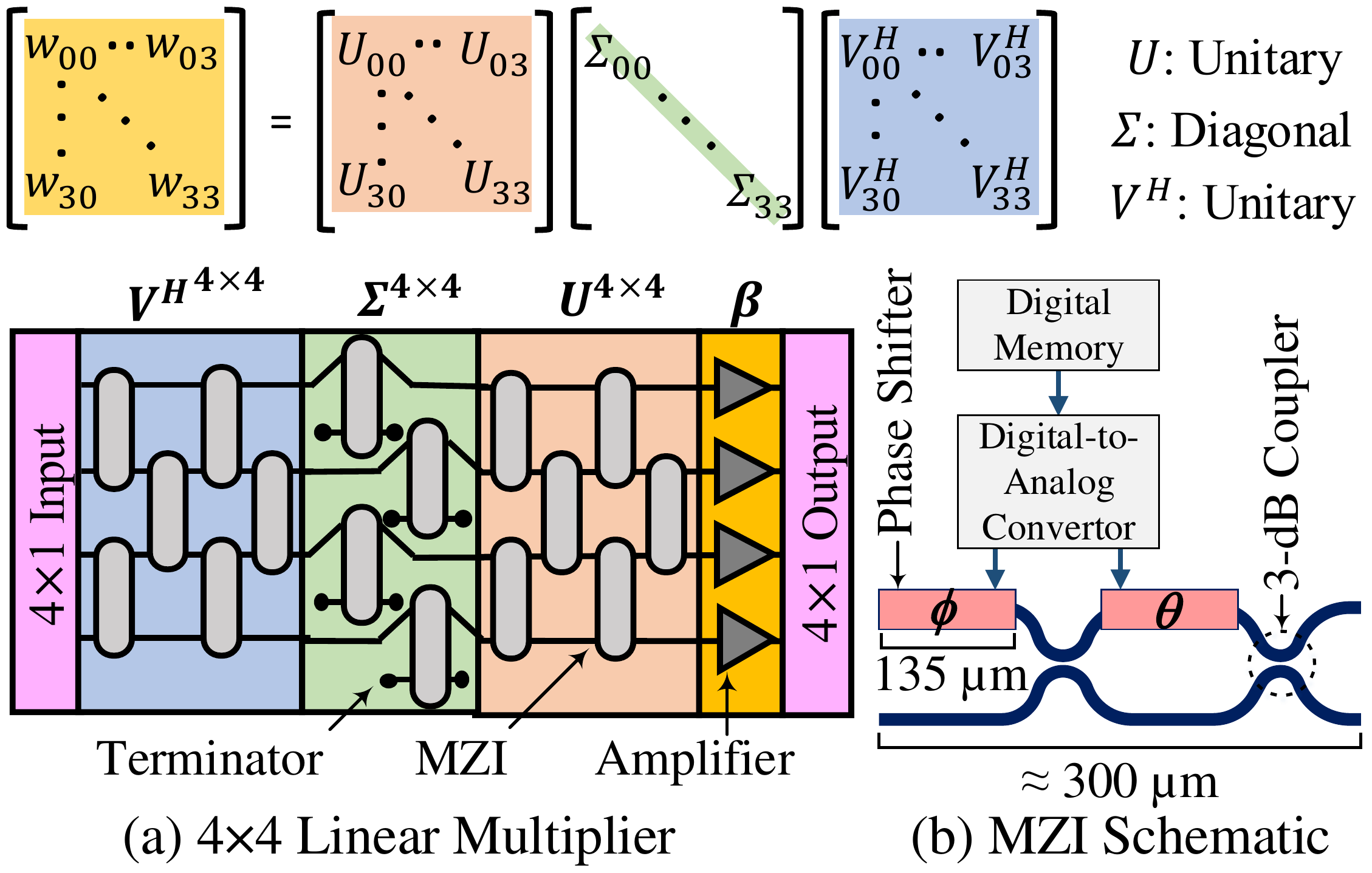}
  \vspace{-0.1in}
  \caption{ (a) A 4$\times$4 linear multiplier realized using MZI arrays, representing the weights of a four-neuron layer connected to another four-neuron layer. (b) An MZI with two phase shifters. Here,  $\phi$ and $\theta$ denote the phase angles adjusted in the phase shifters. The MZI footprint is limited by the length of its phase shifters (MZI dimensions obtained from \cite{shokraneh2020theoretical}).}
\vspace{-0.25in}
  \label{MZIscheme}
 \end{figure}

%Despite many advantages, SC-IPNNs face several challenges that need to be addressed before they can be efficiently deployed. 
As shown in Fig. \ref{MZIscheme}(a), SC-IPNNs use arrays of Mach--Zehnder interferometers (MZIs) as their building block, where the phase angles on each MZI can be adjusted based on the weights in the network and by using training algorithms \cite{fang2019design}. Recent work has shown up to a 70\% accuracy loss in SC-IPNNs due to uncertainties in MZI phase settings \cite{banerjee2021modeling}, and found that such an accuracy loss is mostly due to the uncertainties in MZIs with higher adjusted phase angles \cite{zhu2020countering}\cite{banerjee2021optimizing}. Moreover, SC-IPNNs suffer from large area and static power consumption. In particular, the underlying MZI devices in SC-IPNNs employ lengthy phase shifters (e.g., as long as 135~$\mu$m~\cite{shokraneh2020theoretical}) and consume high static power (e.g., $\approx$25~mW per MZI, dominated by MZI phase shifters \cite{harris2014efficient}). A potential solution to alleviate these problems is to prune SC-IPNNs to reduce the number of bulky components (e.g., phase shifters in MZIs---see Fig. 1(b)) and minimize the phase settings in the network. \par
%phase angles (corresponding to the weight parameters)
%(up to 1000$\times$ that of electronic DNNs, due to the large MZI footprint \cite{shen2016increasing})

While software pruning of weights has shown promising results in electronic implementations of DNNs \cite{han2015deep}, its applicability to SC-IPNN model compaction is significantly limited. This is because of the complex mapping between the weights in the fully connected layers (in software) and the phase angles (in hardware) in SC-IPNNs: i.e., each weight is mapped to multiple phase angles and each phase angle is used to realize multiple weights. Consequently, it is extremely challenging to selectively prune the phase angles that only affect the non-critical (low saliency) weights without deviating the critical weights. Prior efforts on pruning SC-IPNNs using existing techniques have shown that only up to 30\% of the phase angles can be pruned without significant degradation in the accuracy \cite{gu2020towards}. 

%several challenges need to be addressed before high-volume manufacturing of silicon-photonic ICs (and IPNNs, in particular) can become feasible. Photonic components are sensitive to inevitable fabrication process variations in device dimensions and parasitic effects such as distributed scattering at roughness and reflection at discontinuities \cite{bogaerts2013design}. 
 In this paper, we present the first efficient hardware-aware pruning method for SC-IPNNs based on the lottery ticket hypothesis (LTH). This hypothesis states that given any randomly initialized, dense, feed-forward DNN, there exists a sub-network that---when trained from scratch---can match the test accuracy of the original DNN \cite{frankle2018lottery}. Recent work on model compaction of software DNNs has empirically demonstrated the existence of such sub-networks (\textit{winning tickets}), which are 10--20\% of the size of the original network. By leveraging insights from LTH, our pruning method identifies a small subset of phase angles in SC-IPNNs that are critical for maintaining the classification accuracy. As a result, we can reduce the footprint and static power consumption in SC-IPNNs either by removing or power-gating the redundant phase shifters in SC-IPNNs. We consider an SC-IPNN case study with two hidden layers (each with 16 neurons) and show that we can prune up to 89\% of the 1290 phase shifters, and achieve an 86\% reduction in the static power consumption. As we will discuss, pruning of SC-IPNNs with low accuracy loss is a considerably more challenging problem compared to that in electronic implementations of DNNs. To the best of our knowledge, the proposed method is the first to achieve highly sparse SC-IPNNs with less than 5\% accuracy loss. While we focus on SC-IPNNs in this paper, our approach can be easily extended to noncoherent networks. The main contributions of this paper are:
\begin{itemize}
    \item Identifying the challenges associated with pruning SC-IPNNs and the limitations of conventional hardware-unaware software pruning;
    \item Developing the first hardware-aware pruning method based on LTH to generate power- and area-efficient SC-IPNNs;
    \item Exploring the trade-off between the sparsity of the phase angles in a pruned SC-IPNN and its sensitivity to random uncertainties in the (remaining) non-zero phase angles.  
\end{itemize}

The rest of the paper is organized as follows. Section 2 covers the fundamentals of SC-IPNNs and LTH. In Section 3, we highlight the drawbacks of hardware-unaware software pruning in SC-IPNNs. We then describe a hardware-aware magnitude-based (baseline) pruning method and the proposed LTH-based pruning method. Section 4 presents simulation results highlighting the performance of the pruned networks. Finally, we draw conclusions in Section 5.    

\section{Background and Motivation}
%In recent work, several different implementations of integrated photonic neurons (and consequently, IPNNs) have been proposed \cite{sunny2021survey}. For example, in non-coherent IPNNs, the weights are realized by tuning microring resonator banks; however such architectures are prone to inter-channel crosstalk and require several wavelength conversion steps. Therefore, in this paper, we focus on pruning SC-IPNNs. 
%This section presents preliminaries of SC-IPNNs and LTH, and motivates the need for pruning SC-IPNNs. \par

\subsection{Coherent Integrated Photonic Neural Networks}
%The neurons in adjacent linear layers in an MLP are interconnected using weighted edges; these weights are tuned during training to change the effect of each input to the linear layer. The weighted output vector from each layer is then passed through a non-linear activation block (e.g., ReLU, sigmoid).
%Using singular value decomposition (SVD), the weight matrices in the linear layers in multi-layer perceptrons (MLPs) are factorized into two unitary and one diagonal matrix; these matrices are realized using an array of Mach-Zehnder interferometers (MZIs). Each MZI consists of two phase-shifters (PhS) with tunable phase angles and two 50:50 beam splitters (BeS). Prior work shows that $N\timesN$ unitary and diagonal matrices can be represented by tuning the phase angles in an array of $N(N-1)/2$ and $N$ MZIs respectively \cite{clements2016optimal}.  
SC-IPNNs operate with a single wavelength and employ optical phase-change mechanisms in on-chip interferometric devices (i.e., MZIs) to imprint weight parameters onto the electrical field amplitude of optical signals \cite{sunny2021survey}. Using SVD, the weight matrices in linear layers of multi-layer perceptrons can be factorized into two unitary and one diagonal matrices. Using the Clements design \cite{clements2016optimal} and considering Fig.~\ref{MZIscheme}(a), the weight matrix of a linear layer can be realized using three MZI arrays as $W=U\Sigma V^H$, where $V^{H}$ is the Hermitian transpose of $V$. An $N\times N$ unitary and diagonal matrix can be implemented by adjusting the phase angles in an array with $N(N-1)/2$ and $N$ MZIs, respectively. In addition, global optical amplification (layer $\beta$ in Fig. \ref{MZIscheme}(a)) is necessary on each output \cite{connelly2007semiconductor}. The non-linear activation can be performed using opto-electronic units \cite{williamson2019reprogrammable}, not shown in Fig.~\ref{MZIscheme}(a) for the sake of brevity.

Fig.~\ref{MZIscheme}(b) shows a schematic of a 2$\times$2 MZI with two phase shifters (PSes), where $\phi$ and $\theta$ are the phase angles. PSes are used to determine the relative phase difference between the two optical signals traversing the MZI arms and can be implemented using microheaters that work based on the thermo-optic effect in silicon \cite{fang2019design}. In a thermo-optic PS, the temperature-induced phase shift ($\Delta\phi$ or $\Delta\theta$) is proportional to the temperature change ($\Delta T$) based on $\Delta\phi=\left(\frac{2\pi L}{\lambda_0}\right)\cdot \left(\frac{dn}{dT}\right)\cdot \Delta T$.
Here, $L$ is the length of the PS and $\lambda_0$ is the optical wavelength \cite{jacques2019optimization}. Also, $\frac{dn}{dT}\approx 1.8\cdot10^{-4}$ K$^{-1}$ is the thermo-optic coefficient of silicon at $\lambda_{0}=$~1550~nm and temperature $T=$~300~K. The parameter $\Delta T$ can be controlled by applying a DC voltage using a digital-to-analog converter (see Fig. \ref{MZIscheme}(b)). The PS power consumption $P$ is directly proportional to $\Delta T$: $P \propto \Delta T$ \cite{jacques2019optimization}. The MZI also includes two directional couplers with a nominal splitting ratio of 50:50 (3-dB couplers). Based on the two PSes and the two 3-dB couplers, the transfer matrix of a 2$\times$2 MZI multiplier can be easily defined (e.g., see equation 1 in \cite{shokraneh2020theoretical}).
%Here, $G$ denotes the thermal conductance between the heated waveguide and the heat sink, and $A$ denotes the cross-sectional area traversed by the heat flow. 
%\begin{equation}
%    \begin{split}
%       &T_{MZI}(\theta, \phi)=U_{BeS}\cdot U_{PhS}(\theta)\cdot U_{BeS}\cdot U_{PhS}(\phi)\\
%       &=\begin{pmatrix}
%       T_{11} & T_{12} \\
%        T_{21} & T_{22}
%       \end{pmatrix} =  \begin{pmatrix}
%        \frac{e^{i\phi}}{2}(e^{i\theta}-1) & \frac{i}{2}(e^{i\theta}+1) \\
%        \frac{ie^{i\phi}}{2}(e^{i\theta}+1) & -\frac{1}{2}(e^{i\theta}-1) 
%        \end{pmatrix} 
%    \end{split},
%\end{equation}
%where $U_{BeS}$ ($U_{PhS}$) is the BeS (PhS) transfer matrix. \par

Software training of SC-IPNNs can be performed either in a hardware-unaware or in a \textit{photonic} hardware-aware manner. In the hardware-unaware approach, the optimal DNN weight matrices are first obtained using training and are then mapped to different phase angles in the MZIs. In contrast, in \textit{photonic} hardware-aware software training, backpropagation is performed on the phase angles that are adjusted based on the computed gradients. We employ this approach as it offers more control on the phase angles during software training. This is essential for efficient pruning of SC-IPNNs (see Section 3). 

\subsection{Pruning SC-IPNNs: Motivation}
Pruning of neural networks has been the subject of considerable research in recent years. Effective pruning can reduce the infrastructure costs associated with the storage and computation of the weight parameters of enormous DNNs, thereby enabling their deployment in resource-constrained environments. In addition to these improvements, pruning is crucial in SC-IPNNs to improve their area- and power-efficiency, as discussed next. \par

The phase shifters in SC-IPNNs with MZI arrays consume a significant portion of the network area and power. In particular, the size of the constituent thermo-optic PSes in an MZI determines the size of the device (see Fig. 1(b)). For example, the state-of-the-art 2$\times$2 MZI proposed in \cite{shokraneh2020theoretical} is $\approx$~300~$\mu$m long, in which each PS has a length of 135~$\mu$m (i.e., $\approx$~90\% of the length of the MZI considering the two PSes). Moreover, as discussed in Section 2.1, the required phase shift in a PS ($\Delta\phi$) is directly proportional to its length ($L$) and power consumption ($P$): $\Delta\phi\propto L\cdot P$. Even power-efficient PSes can consume up to $\approx$~25~mW DC power for a phase shift of $\pi$ \cite{harris2014efficient}. Note that DC power in PSes is consumed during inferencing to maintain the phase angles. As a result, low accuracy-loss pruning approaches are essential in SC-IPNNs to identify prunable PSes, thereby reducing the network footprint and power consumption. Additionally, as lower phase shifts require lower $\Delta T$, thermal crosstalk between PSes can be minimized by pruning. The problem of explicitly reducing thermal crosstalk is beyond the scope of this paper.   \par

%In addition, photonic waveguides necessitate a minimum bend radius to avoid high optical losses and need to be placed sufficiently apart to avoid evanescent coupling \cite{bogaerts2013design}. These challenges, complemented by the sensitivity of photonic layouts to minute geometric variations, lead to a low integration density in silicon-photonic ICs (compared to electronic ICs at the nanometer regime). 
%a typical silicon waveguide on a 200 mm wafer consists of a 2 $\mu m$ silica lower cladding, 220 nm silicon core and silica upper cladding \cite{wu2020state}. Similarly, ultra-area-efficient BeS can have a footprint of at least 2.4$\times$2.4 $\mu m^2$ \cite{shen2016increasing}. Especially, 
% \textcolor{red}{Should we add uncertaintiy attack surface reduction? Mention also thermal crosstalk.}

\subsection{Lottery Ticket Hypothesis (LTH)}
Recent studies have shown that training a pruned model from scratch is considerably difficult and it often achieves lower accuracy compared to the original (unpruned) model \cite{han2015deep}. Nevertheless, LTH has established that for a given randomly initialized network, we can always find a smaller sub-network that---when trained from scratch---can match the accuracy of the original network within a few training iterations \cite{frankle2018lottery}. These high-performing trainable sub-networks, called \textit{the winning tickets}, can be identified by a modified magnitude-based pruning approach. After the smallest-magnitude weights (below a pre-determined threshold) are pruned, the remaining non-zero parameters are reset back to their original values (before the onset of training). This step is followed by retraining to recover the network accuracy. Experimental results show that the \textit{winning ticket} for a network varies based on the initial weight values \cite{frankle2018lottery}. One possible explanation for this is that the \textit{winning ticket} initialization can potentially land in a region-of-the-loss landscape that enables quick optimization. Model-compaction techniques have shown that stochastic-gradient descent seeks out and trains a sub-network. However, LTH highlights that there exist multiple such sub-networks unique to different initializations (see \cite{frankle2018lottery} for more details on LTH).

%The non-uniqueness of SVD under reflections has been used to realize weight matrices in SC-IPNNs with minimal phase angles in \cite{banerjee2021optimizing}. Using a simulated annealing-based optimal reflection search method, the tuned phase angles (and hence the static power consumption) can be reduced by up to 15.3\% without any accuracy loss. While this improvement is significant, the reduction in phase angles will likey get amortized over large SC-IPNN architectures. 

% \subsection{Lottery Ticket Hypothesis (LTH)}

\section{Enabling Efficient Pruning in SC-IPNNs}

\subsection{Challenges in Pruning SC-IPNNs}\label{challenges}
Hardware-unaware software pruning methods in DNNs aim at obtaining a sparse weight matrix \cite{han2015deep}. A binary mask $M^k$ is maintained for each DNN layer $L^k$. An element of the mask, say $M^{k}_{i,j}$, is 0 (1) iff the corresponding weight $L^{k}_{i,j}$ is zero (non-zero). In each pruning iteration, a fraction of the non-zero weights---typically those with a smaller magnitude---in each layer is clamped to zero, and the corresponding mask elements are updated. During backpropagation in retraining, the gradient of each weight is multiplied with its respective mask element, ensuring that the zero weights in each layer are not updated. 
\begin{figure}[t]
 \centering
  \includegraphics[width=.48\textwidth]{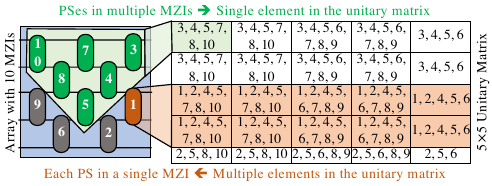}
  \vspace{-0.2in}
  \caption{An example of the bidirectional many-to-one association (BMA) between the elements of the weight matrix and the MZI array for a 5$\times$5 unitary matrix. The numbers in each cell of the unitary matrix denote the MZIs that affect the corresponding matrix element.}
\vspace{-0.2in}
  \label{bidi}
 \end{figure}

\begin{figure}[t]
 \centering
  \includegraphics[width=.48\textwidth]{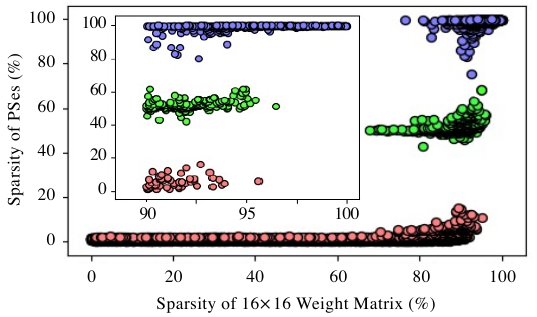}
  \vspace{-0.2in}
  \caption{Comparison between the sparsity of 10000 randomly generated 16$\times$16 unitary matrices and the sparsity of the PSes (when phase angles are zero) in the respective MZI devices. Inset: Comparison of the matrix sparsity with the sparsity of PSes for 1000 highly sparse (matrix sparsity $>$90\%) 16$\times$16 unitary matrices. The red, green, and blue clusters indicate unitary matrices that are mapped to MZI arrays with low, medium, and high sparsity in the PSes, respectively.}
\vspace{-0.15in}
  \label{sparsity}
 \end{figure}

Unfortunately, there are several problems with applying such \textcolor{black}{DNN} software pruning techniques to SC-IPNNs. In particular, each element of the weight matrix of a linear layer in SC-IPNNs is mapped to multiple phase angles, and each phase angle in an MZI array affects multiple elements of the weight matrix. Fig. \ref{bidi} shows an example of this \textit{bidirectional many-to-one association (BMA)} between a 5$\times$5 unitary matrix and its corresponding MZI array. Due to this BMA, if a phase angle in an MZI is updated to prune a non-critical (low-magnitude) weight, it can also affect another potentially critical weight, thereby leading to significant accuracy losses. Moreover, because of BMA, a sparse weight matrix may not necessarily lead to \textit{sparsity in the PSes of the corresponding MZI devices, i.e., when one or both of the phase angles ($\phi$ and $\theta$) are zero and the PSes can be then removed or power-gated}. Fig. \ref{sparsity} compares the sparsity of 10000 randomly generated 16$\times$16 unitary matrices with the sparsity of their mapped PSes in the corresponding MZIs. Observe that a highly sparse ($>$90\%) weight matrix does not always lead to sparsity in PSes. While software pruning focuses on a sparse weight matrix, SC-IPNN model compaction should minimize and prune MZI phase angles to reduce area overhead and power consumption (see Section 2.2). The discrepancy between these two objectives indicates the ineffectiveness of software pruning in SC-IPNNs and the critical need for hardware-aware pruning. \par

By leveraging hardware-unaware software pruning, where the weight matrices are first pruned in software and then mapped to MZI arrays, \cite{gu2020towards} showed that no more than 30\% of the phase angles can be pruned without a significant accuracy loss ($\approx$10\%) in SC-IPNNs. To address this, \cite{gu2020towards} proposed a pruning-friendly \textit{non-SVD-based} C-IPNN architecture that leverages block-circulant matrix representation and performs matrix-vector multiplication using optical fast Fourier transform (FFT). However, even this alternative architecture could achieve a sparsity of only up to 45\%. Unlike in SC-IPNNs, hardware-unaware software pruning is applicable to noncoherent IPNNs. In \cite{zhang2021implementation}, pruned noncoherent IPNNs demonstrate an accuracy of up to 93.49\% on the MNIST dataset. Using layer-wise pruning and weight clustering, \cite{sunny2021sonic} obtains a sparsity of up to 50\% in a noncoherent neural network inference accelerator.

\subsection{Hardware-Aware Pruning in SC-IPNNs}
As discussed in Section \ref{challenges}, \textcolor{black}{contrary to electronic DNNs, }pruning in SC-IPNNs must be hardware-aware to ensure sparsity in phase angles. However, as we will show in Section 4, the nonlinear dependence between the phase angles and the weights in SC-IPNNs---e.g., weights with large magnitude can be mapped to smaller phase angles---makes it challenging to identify non-critical weights that can be safely pruned, even using hardware-aware techniques. Consequently, simply pruning phase angles based on their magnitude can be ineffective. In this section, we first present a baseline approach where we apply conventional magnitude-based pruning in which we consider the magnitude of the adjusted phase angles. Next, we present the LTH-based pruning technique and show that it can prune a significant fraction of phase angles with a negligible accuracy loss. \par

\subsubsection{Baseline Method: Phase-Angle-based Magnitude Pruning}\label{baseline_sec}
In magnitude pruning, all the weights in a layer having magnitudes smaller than a threshold are set to zero. Next, in post-pruning, the network is retrained to recover the lost accuracy. However, the pruned weights are kept at zero during the retraining. There are several ways to determine the threshold for a layer. For example, in magnitude pruning based on the mean (standard deviation), the threshold is considered to be a factor---say $\alpha$---of the mean (standard deviation) of the non-zero weights in a layer. Magnitude pruning can be performed either in one shot or in an iterative manner. In one-shot pruning, all the weights below a threshold are pruned in a single step after which retraining (a.k.a. fine-tuning) is performed. In iterative pruning, weights are gradually pruned over multiple steps with each step followed by few rounds of fine-tuning. While extending magnitude pruning to SC-IPNNs, we implement both the one-shot and the iterative approaches. We consider \textit{photonic} hardware-aware software training, and therefore, during backpropagation, gradients are calculated for each phase angle (and not layer edge weights). Consequently, the binary masks used to suppress the gradients for the pruned phase angles are also maintained for the PSes in the MZI array corresponding to the weight matrix of each linear layer. 
 
\subsubsection{Proposed Method: Using LTH to Prune SC-IPNNs}
\begin{figure}[t]
 \centering
  \includegraphics[scale=2]{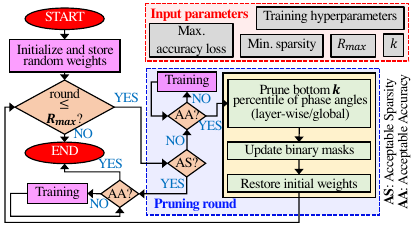}
  \vspace{-0.2in}
  \caption{An overview of the proposed LTH-based hardware-aware pruning method for SC-IPNNs. The input parameters are user-defined.}
\vspace{-0.25in}
  \label{blockdiag}
 \end{figure}

Pruning based on LTH is similar to magnitude pruning, but with a significant difference: after the weights are pruned, the remaining non-zero weights are set back to their initial values, which were stored before training began for the first time. This step is followed by retraining to recover the lost accuracy. Fig. \ref{blockdiag} presents an overview of the proposed LTH-based pruning for SC-IPNNs. The inputs are the hyperparameters (learning rate and epochs) for the retraining step, the maximum acceptable accuracy loss, minimum sparsity, maximum number of pruning rounds ($R_{max}$),  and the pruning rate ($k$). As a pre-processing step, we initialize the weights and store their values in a database. In each round, we check whether the network sparsity is acceptable, in which case we exit the pruning round and retrain to achieve acceptable classification accuracy. Otherwise, we enter the round and retrain the network. Note that the training hyperparameters should be adjusted in each round to obtain a sufficiently high accuracy ($<$~5\% accuracy loss) before the phase angles are pruned. After retraining, the bottom $k$ percentile of the phase angles with small magnitude are pruned and the respective binary masks are updated. The $k$-percentile pruning step can be performed in two ways: in layer-wise LTH pruning, the bottom $k$ percentile weights in each tier are pruned, whereas, in global pruning, the bottom $k$ percentile weights in the entire SC-IPNN are pruned. After pruning, we set the non-zero weights back to their initial values before proceeding to the next round. We use LTH to identify the best-performing \textit{winning ticket} iteratively over multiple rounds as it has been shown to yield more sparse subnetworks compared to the one-shot (single-round) approach \cite{frankle2018lottery}.  \par

 Due to the unique challenges associated with the pruning of SC-IPNNs (see Section 3.1), we encounter a significant loss in accuracy after each pruning iteration. However, by resetting the non-zero phase angles to their initial values after each iteration, our approach gradually identifies the best solution (with acceptable sparsity and accuracy) out of the \textit{winning tickets} in each iteration. In accordance with LTH, this ensures that the pruned models obtained using our method can recover the accuracy loss (due to pruning) more effectively compared to a pruned model of similar sparsity obtained using the baseline method. As a result, we are able to achieve significantly higher sparsity with a negligible accuracy loss, as discussed in the next section. Note that the proposed pruning method should be performed offline and only once per SC-IPNN design.

 \section{Simulation and Evaluation Results}
We use a fully connected feedforward SC-IPNN with two hidden layers (i.e., 32 neurons and 1290 PSes) to demonstrate the performance of the baseline and the proposed LTH-based pruning methods. Each linear layer is implemented using the Clements design \cite{clements2016optimal} and is followed by a nonlinear Softplus function. To model intensity measurement, a modulus squared nonlinearity is applied after the output layer. This is followed by a final LogSoftMax layer to obtain a probability distribution. We use a cross-entropy loss function \cite{cover2006elements} to train the SC-IPNN on the MNIST dataset \cite{lecun1998mnist}. Similar to \cite{banerjee2021modeling}, we use shifted FFTs to convert each 28$\times$28 MNIST image to a 16-dimensional complex feature vector. The nominal inferencing accuracy (on the test dataset) of the unpruned SC-IPNN is 93.86\%. While we use this case-study for our simulations, the proposed method is agnostic to the network (number and size of linear layers, non-linear activation, and loss functions). 
\begin{figure}[t]
 \centering
   \subfigure[Baseline method]
   {\includegraphics[scale=1.05]{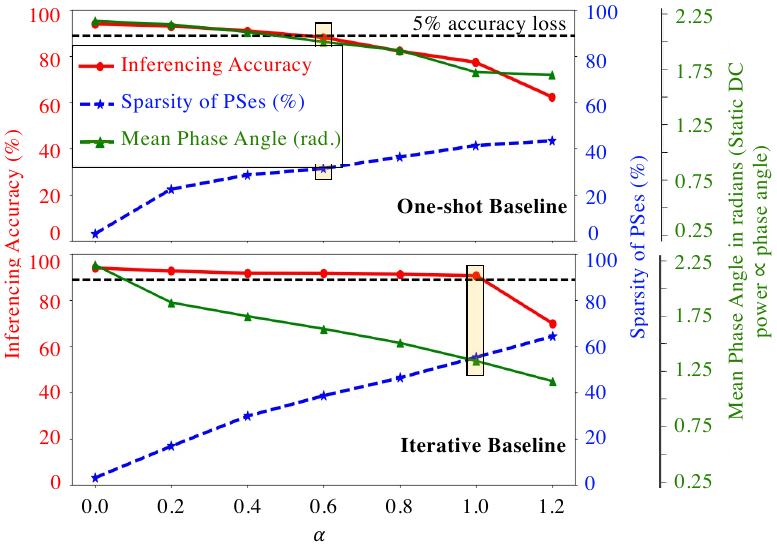}\label{baseline}}\vspace{-0.1in} 
  \subfigure[Proposed LTH-based pruning method]{
  \hspace{0.1em}
  \includegraphics[scale=1.05]{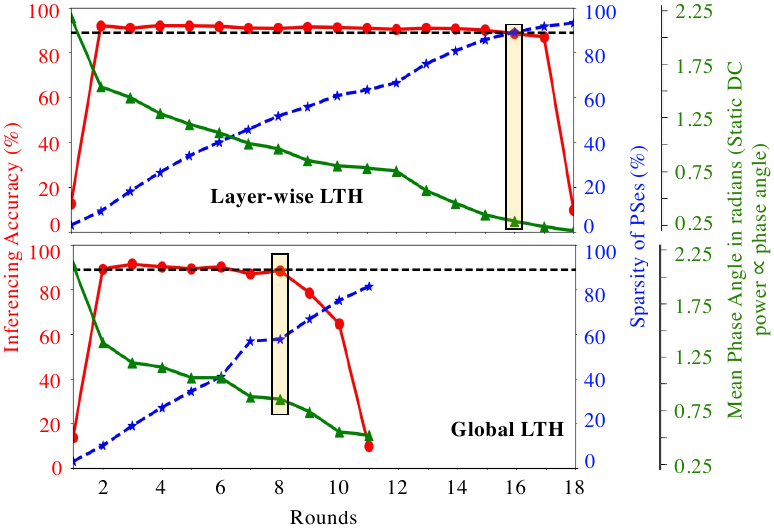} \label{ltpplots}
  } \vspace{-0.1in}
  \caption{Fine-tuned accuracy, PS sparsity, and mean phase angle for (a) one-shot phase-angle magnitude-based (baseline) pruning (top) and iterative baseline pruning (bottom) with different values of $\alpha$; and (b) different rounds of layer-wise (top) and global LTH-based pruning (bottom). The black-dashed lines show a 5\% accuracy loss and the yellow rectangles highlight the best-performing models (maximum sparsity with accuracy loss $<$5\%) for each pruning method.}
 \vspace{-0.15in}
 \end{figure}

 \subsection{Pruning Analysis for SC-IPNNs} 
Fig. \ref{baseline} shows the simulation results for the one-shot (top) and iterative (bottom) phase-angle magnitude-based pruning with the baseline method. In each case, we consider standard-deviation-based pruning (and not the mean-based one) to calculate the threshold as it takes the distribution of the phase angles in each layer into account. The threshold is given by $\alpha\cdot\sigma_{layer}$. Here, $\alpha$ is a user-defined constant (see Section \ref{baseline_sec}) and $\sigma_{layer}$ denotes the standard deviation of the non-zero phase angles in a layer. We consider both one-shot and iterative approaches as in a few cases (e.g., for $\alpha=0.2$), the former performs better. For iterative pruning, we approach this threshold in incremental steps of $\alpha\cdot\sigma_{layer}/$10. As can be seen, in both cases the overall sparsity of the PSes increases with $\alpha$. Also, the mean phase angle---averaged over the 1290 PSes to which the weight parameters are mapped---and hence the static DC power (see Section 2.2) decrease as $\alpha$ increases. However, the inferencing accuracy drops significantly for the one-shot pruning as $\alpha$ increases. In contrast, in iterative pruning, the accuracy loss is less than 5\% up to $\alpha=$1 (see the black-dashed line in Fig. \ref{baseline}). This is because of the gradual pruning and fine-tuning in iterative pruning, compared to the drastic pruning and few fine-tuning iterations in the one-shot case. If one allows for an accuracy loss of 5\%, up to 55\% (31\%) PSes can be pruned using the iterative (one-shot) approach. \par
\begin{figure*}[t]
  \centering
  \subfigure[]{
\includegraphics[width=.32\textwidth]{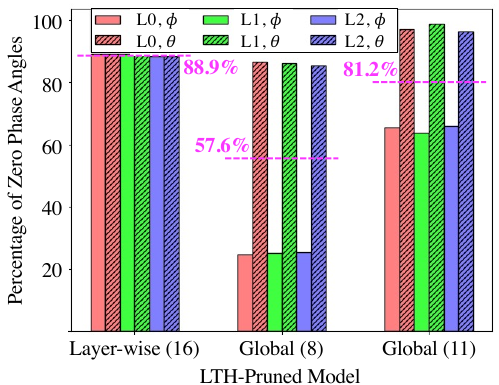}
}%
\hspace{-1em}
\subfigure[]{
\includegraphics[width=.33\textwidth]{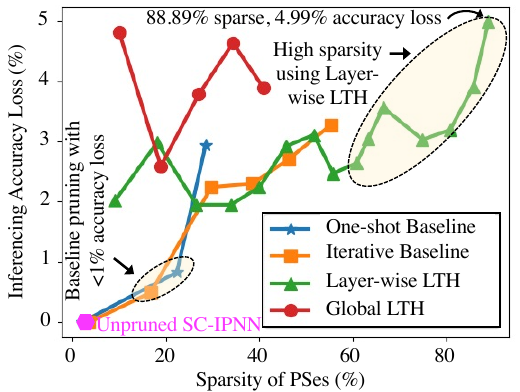}
}%
\hspace{-1em}
\subfigure[]{
\includegraphics[width=.33\textwidth]{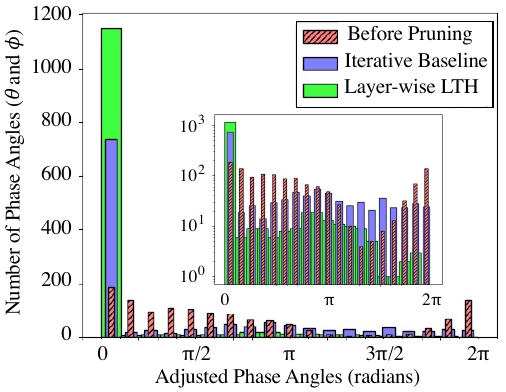}
}%
\vspace{-0.2in}
  \caption{(a) Phase-angle sparsity ($\phi$ and $\theta$) in the three layers (L0, L1, and L2) of different LTH-pruned models. The x-axis shows the variant of LTH pruning used with the number of rounds in parentheses. The magenta-dashed lines indicate the mean sparsity over all the layers. (b) Comparison between the accuracy loss and PS sparsity in pruned models obtained using different methods. (c) Histogram distribution of the phase angles in the SC-IPNN with baseline and LTH-based pruning (inset shows the same plot with a logarithmic scale on the y-axis).}
 \vspace{-0.15in}
  \label{combo}
\end{figure*}

For the LTH-based pruning method, we found that the performance is better when we start with a low pruning rate ($k$) for the first few rounds, before aggressively pruning (high $k$) in the final rounds. Accordingly, we consider a pruning rate of $k=$10\% for the first ten rounds and then increase it to $k=$25\% for the remaining rounds. To show results over several rounds, we do not constrain the maximum accuracy loss, minimum sparsity, and $R_{max}$ parameters, and tune the training hyperparameters to maximize the accuracy. Fig. \ref{ltpplots} shows the results for the proposed layer-wise (top) and global (bottom) LTH-based pruning. Recall that in layer-wise (global) LTH-based pruning, the non-zero phase angles in the bottom $k$ percentile of each layer (the entire SC-IPNN) are pruned. In both cases, the inferencing accuracy is significantly low in the first round as it is computed before the network is trained (see Fig. \ref{blockdiag}). The accuracy also drops sharply in the final round (round 18 for layer-wise and 11 for global) as the loss function explodes when a large fraction of phase angles are pruned, thereby leading to erroneous gradient propagation. As expected, the PS sparsity increases and the mean phase angle as well as the static DC power consumption (see Section 2.2) decrease with more rounds of pruning. Yet, the accuracy loss is smaller than 5\% (above the black-dashed line in Fig. \ref{ltpplots}) for many rounds, especially for the layer-wise pruning. In fact, up to 89\% of the PSes can be pruned with 86\% reduction in the mean phase angles and static power consumption using 16 rounds of layer-wise LTH-based pruning. Similarly, up to 57\% of phase angles can be pruned using the global LTH pruning. \par

The global LTH-based pruning performs worse compared to the layer-wise pruning as it is biased towards layers with smaller phase angles, i.e., it is possible that most phase angles in such layers are pruned away in the initial rounds. This is highlighted in Fig. \ref{combo}(a) where we show the sparsity of phase angles in the three (one input and two hidden) layers of layer-wise and global LTH-pruned models. In the global model with the best trade-off between accuracy and sparsity (eight rounds of pruning, 88.9\% accuracy, and 57.6\% mean sparsity), the percentage of pruned (i.e., zero) $\theta$ phase angles is considerably higher than $\phi$. In the global model with maximum sparsity (11 rounds of pruning and 81.2\% mean sparsity), up to 98.8\% of $\theta$ phase angles are pruned. Extreme sparsity in certain layers can potentially hinder training and lead to exploding loss. In contrast, layer-wise pruning (see Fig. \ref{combo}(a)) ensures that the sparsity of phase angles is uniform across the different layers. \par

When considering both high sparsity and low accuracy loss, the LTH-based pruning outperforms hardware-aware magnitude pruning (baseline). Fig. \ref{combo}(b) compares the accuracy loss and sparsity of pruned SC-IPNNs obtained using different methods. Here, we consider only those models where the accuracy loss due to pruning is less than 5\%. The unpruned SC-IPNN has 3.03\% sparsity based on the initialization (see the magenta data point). From Fig.~\ref{combo}(b) it is clear that only layer-wise LTH pruning can offer a sparsity greater than 60\%. When very low accuracy loss ($<$1\%) is acceptable after pruning, the hardware-aware magnitude pruning (baseline) can be considered, which achieves a maximum sparsity of 22\% under this constraint. Fig. \ref{combo}(c) compares the histograms of the phase angles of the best-performing models obtained using the baseline and the LTH-based pruning with that of the unpruned SC-IPNN. Layer-wise LTH-based pruning not only achieves high sparsity with negligible accuracy loss but minimizes the phase angles, resulting in significant savings in static power consumption (phase angle and static power are linearly proportional).  

%In summary, we find that layer-wise LTH-based pruning is most likely to be the best performing hardware-aware pruning technique for SC-IPNNs. In special cases where the classification accuracy is crucial, the iterative baseline method may be considered.  
\subsection{Noise Sensitivity of Pruned SC-IPNNs}
%Due to fabrication process variations and thermal crosstalk, the phase angles in the PhS can deviate from their tuned values. Such deviations lead to faulty matrix multiplication in the linear layers, thereby imposing classification accuracy loss in SC-IPNNs. Prior studies have mentioned that an error of up to 0.21 radians can be expected in the phase angles even for mature fabrication processes \cite{flamini2017benchmarking}. Such random deviations in the phase angles can lead to up to 70\% degradation in the SC-IPNN classification accuracy. \par   

\begin{figure}[t]
 \centering
  \includegraphics[width=.35\textwidth]{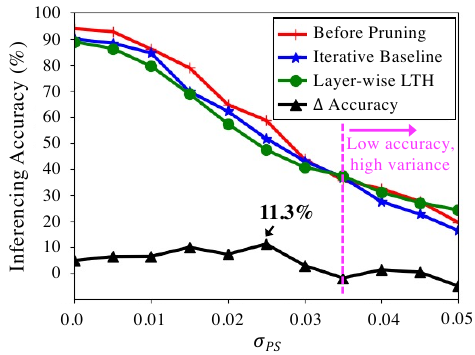}
  \vspace{-0.1in}
  \caption{Inferencing accuracy of the unpruned SC-IPNN and the best-performing pruned models obtained using the baseline and the LTH-based pruning methods. $\Delta$Accuracy denotes the difference in the accuracy between the unpruned and the LTH-pruned models.}
\vspace{-0.2in}
  \label{accloss}
 \end{figure}

As the redundant parameters in a DNN are gradually discarded during model compaction, the pruned DNN becomes more sensitive to uncertainties in the (few) remaining parameters. This is indeed critical for sparse SC-IPNNs because even overparameterized and unpruned SC-IPNNs are sensitive to uncertainties, especially those in the adjusted phase angles in the network \cite{banerjee2021modeling}. Using the SC-IPNN case study in this paper, we consider 1000 Monte Carlo iterations and inject random uncertainties in the phase angles of the unpruned SC-IPNN and those of the best-performing models obtained from the iterative baseline and the layer-wise LTH-based pruning methods. In each iteration, the uncertainties are sampled from a zero-mean Gaussian distribution with a standard deviation of $\sigma_{PS}\cdot\pi$. Fig. \ref{accloss} shows the mean inferencing accuracy---over 1000 Monte Carlo iterations---for the three models for different values of $\sigma_{PS}$. We observe that while all three models are sensitive to uncertainties, the degradation in accuracy is slightly higher in sparse networks. The black-triangle line in Fig. \ref{accloss} shows that the difference in the accuracy of the unpruned and the LTH-pruned model can be up to 11.3\%. Note that in the absence of phase uncertainties ($\sigma_{PS}=0$), the accuracy of the unpruned network is only 5\% greater than the LTH-pruned model. Under uncertainties, the baseline-pruned model has slightly higher accuracy (0.4\% on average) than the LTH-pruned model as it is less sparse. Mitigating uncertainties in SC-IPNNs often imposes high power-consumption overhead (e.g., when using tuning mechanisms \cite{cheng2020silicon,fang2019design}). Fortunately, LTH-based pruning offers significant savings in power consumption, facilitating the deployment of uncertainty-mitigation methods in SC-IPNNs.

%Notice that for higher $\sigma_{PS}$ ($>$~0.35), the accuracies are close and the pruned models outperform the unpruned model in some cases. However, in the presence of such catastrophic uncertainties, all the models are essentially untrained (random guessing),therefore, the observations are not meaningful.   

%\vspace{-1em}
\section{Conclusion}
Pruning is challenging in SC-IPNNs because of the bidirectional many-to-one association between the edge weights of the linear layer and the phase angles. This paper is the first effort at hardware-aware pruning in SC-IPNNs to minimize their area overhead and static power consumption. We have presented two hardware-aware pruning methods, including a conventional magnitude-pruning-based approach for moderate sparsity (up to 22\%) and ultra-low accuracy loss ($<$1\%), and a novel pruning method based on the lottery ticket hypothesis to achieve ultra-high sparsity (up to 89\%) with an acceptable accuracy loss ($<$5\%). The insights derived from this paper pave the way for enabling advanced hardware-software-assisted design-optimization solutions for realizing compact and energy-efficient integrated photonic neural networks. 

%Our comprehensive analysis identifies the challenges in SC-IPNN pruning, proposes a novel low-cost pruning technique, and highlights the reliability concerns associated with aggressive compaction in SC-IPNNs.
%\vspace{-0.3em}

%We also show that while model compaction reduces the resource requirements of SC-IPNNs, it affects their inherent resilience to uncertainties (e.g., those in the phase angles). 

%\section{Acknowledgement}

\bibliographystyle{IEEEtran}
\bibliography{main}
% ACM needs 'a single self-contained file'!
%
\end{document}